\pdfoutput=1 
\documentclass[final,5p,times,twocolumn]{elsarticle}

\usepackage{lineno}
\usepackage{hyperref}
\usepackage{xspace}
\hypersetup{colorlinks=true,%
           linkcolor=black,%
           anchorcolor=black,%
           citecolor=black,%
           filecolor=black,%
           menucolor=black,%
           urlcolor=black}
\usepackage{subfigure}
\usepackage{sidecap}   
\usepackage{caption}
\usepackage{amsmath, amsthm, amssymb}
\usepackage{siunitx}
\usepackage{comment}

\newcommand{\neqcm}{\ensuremath{\mathrm{n}_{\mathrm{eq}}/\mathrm{cm}^2}}

\biboptions{comma,square,sort&compress}

\graphicspath{{./images/}}

\begin{document}
\begin{frontmatter}
\title{Thin n-in-p planar pixel modules for the ATLAS upgrade at HL-LHC}
\author[label1]{N.~Savic}
\ead{Natascha.Savic@mpp.mpg.de}
\author[label1]{L.~Bergbreiter}
\author[label1]{J.~Breuer}
\author[label1]{A.~La Rosa}
\author[label1]{A.~Macchiolo}
\author[label1]{R.~Nisius}
\author[label1]{S.~Terzo}

\address[label1]{Max-Planck-Institut f\"ur Physik (Werner-Heisenberg-Institut),
  F\"ohringer Ring 6, D-80805 M\"unchen, Germany}
%
\begin{abstract}
The ATLAS experiment will undergo a major upgrade of the tracker system in view of the high luminosity phase of the LHC (HL-LHC) foreseen to start around 2025. Thin planar pixel modules are promising candidates to instrument the new pixel system, thanks to the reduced contribution to the material budget and their high charge collection efficiency after irradiation. New designs of the pixel cells, with an optimized biasing structure, have been implemented in n-in-p planar pixel productions with sensor thicknesses of \SI{270}{\micro{}}m. Using beam tests, the gain in hit efficiency is investigated as a function of the received irradiation fluence. The outlook for future thin planar pixel sensor productions will be discussed, with a focus on thin sensors with a thickness of 100 and  \SI{150}{\micro{}}m and a novel design with the optimized biasing structure and small pixel cells (50x50 and 25x100 \SI{}{\micro{m}^2}). These dimensions are foreseen for the new ATLAS read-out chip in 65 nm CMOS technology and the fine segmentation will represent a challenge for the tracking in the forward region of the pixel system at HL-LHC. To predict the performance of 50x50 \SI{}{\micro{m}^2} pixels at high $\eta$, FE-I4 compatible planar pixel sensors have been studied before and after irradiation in beam tests at high incidence angle with respect to the short pixel direction. Results on cluster shapes, charge collection- and hit efficiency will be shown.
\end{abstract}
%
%
\end{frontmatter}
\section{Introduction}
\begin{figure}[h]
	\begin{center}
		\includegraphics[width=0.3\textwidth]{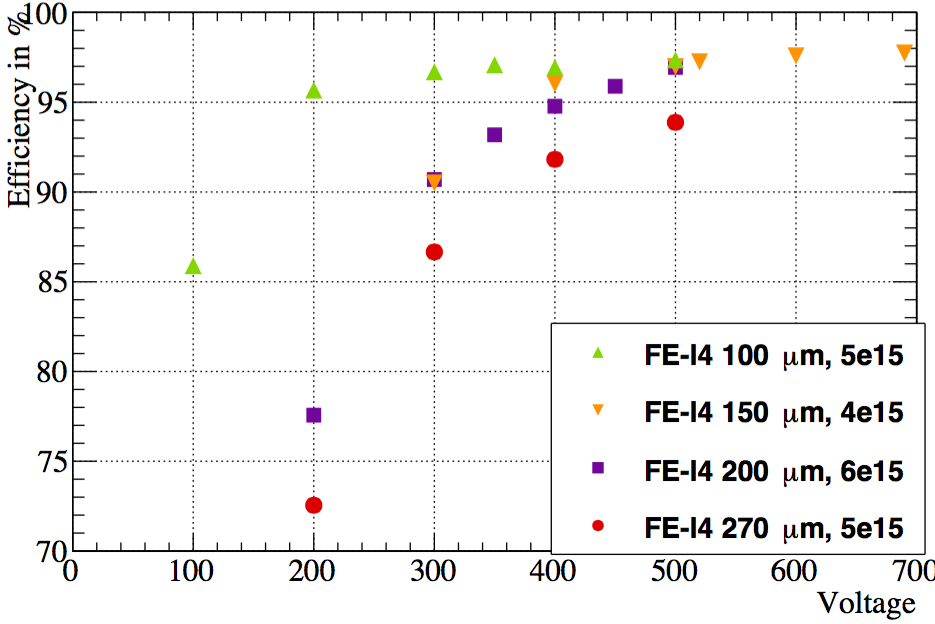}
		\caption[]{Comparison of hit efficiencies of FE-I4 modules with sensor thicknesses between 100 and 270 \SI{}{\micro{m}} at an irradiation fluence of around \SI{5d15}{\neqcm{}}.}
		\label{effthick4}
	\end{center}
\end{figure}

One of the main challenges for the innermost tracking detectors after the upgrade to the HL-LHC with an instantaneous luminosity of up to \SI{5d-34}{\cm^{-2}s^{-1}} will be the exposure to high radiation. The ATLAS pixel system will be prospectively exposed to particle fluences up to \SI{2d16}{\neqcm{}} (1 MeV neutron equivalent)~\cite{atlasUp, Lumi}. To maximize the hit efficiency and reduce the leakage current and power dissipation after irradiation, thin sensors are being developed. Sensors with a thickness of 100 and 150 \SI{}{\micro{m}} were found to reach the same hit efficiency as thicker sensors already at a bias voltage of 300 V shown in Fig~\ref{effthick4}. 
\\A highest hit efficiency of around 97\% was obtained for perpendicular incident tracks at a fluence of \SI{5d15}{\neqcm{}}, the expected fluence for the second layer at HL-LHC~\cite{terzo}. The main inefficiencies are caused by the bias dot and the bias rail, as described in~\cite{anna} for fluences up to \SI{3d15}{\neqcm{}}. In this paper, different designs of n-in-p planar hybrid pixel modules are investigated at the expected fluence of the second layer. Alternative biasing structures were implemented in a CiS sensor production with \SI{270}{\micro{m}} thickness and compared to the standard design. To cope with the highest occupancy at HL-LHC, smaller pixel dimensions with respect to the ones presently implemented in the FE-I3 chip (50x400 \SI{}{\micro{m}^2}) and the FE-I4 chip (50x250 \SI{}{\micro{m}^2}), developed for the ATLAS Insertable B-Layer (IBL), are mandatory~\cite{ibl}. The new read-out chip for the ATLAS pixel systems at HL-LHC is being developed by the CERN RD53 Collaboration with a pixel cell of 50x50 \SI{}{\micro{m}^2} in the 65 nm CMOS technology and is expected to be ready at the beginning of 2017~\cite{rd53, chip}. Sensors compatible with this chip have been implemented in a recent MPG-HLL pixel production with a thickness of 100 and 150 \SI{}{\micro{m}}. First results on the electrical characterization of these devices will be shown.
\section{Optimization of the pixel cell design}
\subsection{Test beam analysis of different pixel cell designs}
\begin{figure}[h]
	\begin{subfigure}[Standard design]
		{\includegraphics[width=0.378\textwidth]{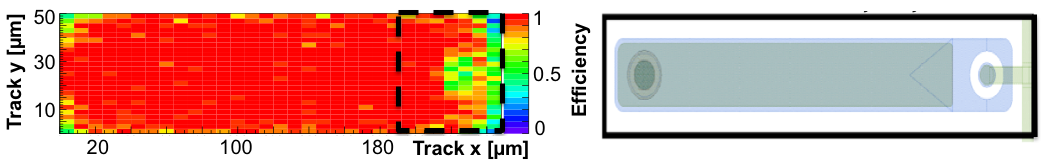}
		}
	\end{subfigure}
	\begin{subfigure}[Modified design]
		{\includegraphics[width=0.378\textwidth]{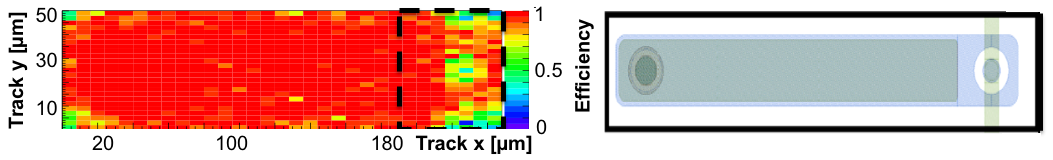}
		}
	\end{subfigure}
	\begin{subfigure}[Common punch-through design]
		{\includegraphics[width=0.48\textwidth]{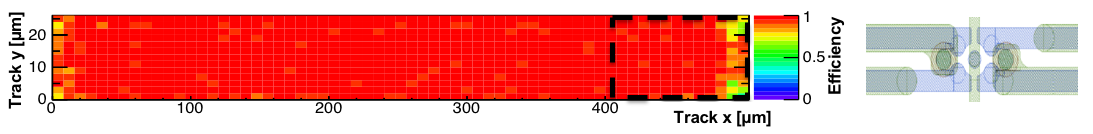}
		}
	\end{subfigure}
	\caption[]{Hit efficiency for (a) the standard punch-through design and (b) the modified individual biasing structure where the bias rail is running over the bias dot after an irradiation fluence of \SI{5d15}{\neqcm{}}. In addition the hit efficiency map of a pixel cell with the common punch-through design with modified dimensions to 25x500 \SI{}{\micro{m}^2} is illustrated in (c). Cut-offs of a 50x50 \SI{}{\micro{m}^2} pixel cell inside the 50x250 \SI{}{\micro{m}^2} pixel cell (a and b) and a 25x100 \SI{}{\micro{m}^2} pixel cell inside the 25x500 \SI{}{\micro{m}^2} pixel cell (c) are outlined. The modules were operated at 500 V.}
	\label{pts}
\end{figure}

In the present design of the ATLAS pixel sensors it is possible to bias the pixel via the punch-through mechanism with an n$^+$ implant dot implemented in the pixel cell, connected to the bias ring through an aluminium rail. It has been observed that these structures introduce a loss of efficiency after irradiation~\cite{anna, terzo}. To reduce this effect, an optimization has been carried out, comparing the performance of the different designs shown in Fig.~\ref{pts}. These were implemented in two FE-I4 compatible sensors in a CiS n-in-p production on 6" wafers with a thickness of 270 \SI{}{\micro{m}}. The sensors were irradiated to a fluence of \SI{5d15}{\neqcm{}} and then studied with a beam test at CERN SPS. The hit efficiency was determined performing track reconstruction with EUTelescope software~\cite{eutelescope}. The systematic uncertainty associated to the efficiency measurements is 0.3\%, as estimated in~\cite{jens}. Also at this higher fluence the relative performance remained similar to the one observed at \SI{3d15}{\neqcm{}}: the bias rail superimposed to the pixel implant as in Fig.~\ref{pts}b yields a higher hit efficiency with respect to the standard geometry. A still better hit efficiency was found for the common punch-through dot, placed externally to the pixel implant and serving four neighbouring pixels, Fig.~\ref{pts}c. This geometry was implemented in a 25x500 \SI{}{\micro{m}^2} pixel cell, still compatible with the FE-I4 chip.
\\At a bias voltage of 500 V a hit efficiency of 93.9\% was obtained for the standard design where the new arrangement of the bias rail in the modified design in Fig.~\ref{pts}b improved the hit efficiency to 94.6\%. The common punch-through yields even 96.0\%, while it increases to 98.0\% at 800 V. These values are about 3\% lower than what was obtained at a fluence of \SI{3d15}{\neqcm{}}. The smaller pixel dimensions of the future pixel read-out chips are indicated in the pixel cells in Fig.~\ref{pts} using a 50x50 (a and b) and 25x100 \SI{}{\micro{m}^2} (c) pixel cell. A smaller pixel with the current designs would show even lower efficiencies. Therefore, further optimization of the biasing structures is mandatory. It is planned to repeat the hit efficiency measurements with sensors in a fluence range up to \SI{d16}{\neqcm{}} to confirm the better performance observed within the new biasing design. 
\subsection{Estimation of hit efficiency for a 25x100 \SI{}{\micro{m}^2} pixel cell}
Since modules with small pixel cells are not available yet, the hit efficiency for a 25x100 \SI{}{\micro{m}^2} pixel cell at an irradiation fluence of \SI{3d15}{\neqcm{}} was estimated based on the existing prototype of the 25x500 \SI{}{\micro{m}^2} pixel cell with the new common punch-through design. Since inefficiencies appear at the edges of the pixel caused by charge sharing as well as by the punch-through structure, the hit efficiencies in the first 40 \SI{}{\micro{m}} and in the last 60 \SI{}{\micro{m}} were combined to estimate the efficiency for an effective pixel cell of 25x100 \SI{}{\micro{m}^2}. A value of 95.5\% was obtained and is indeed lower compared to the 25x500 \SI{}{\micro{m}^2} pixel cell, but only a bit lower compared to the hit efficiency of 96.5\% obtained with the standard implementation of the punch-through in a 50x250 \SI{}{\micro{m}^2} pixel cell.
\begin{figure}[h]
	\begin{center}
		\includegraphics[width=0.47\textwidth]{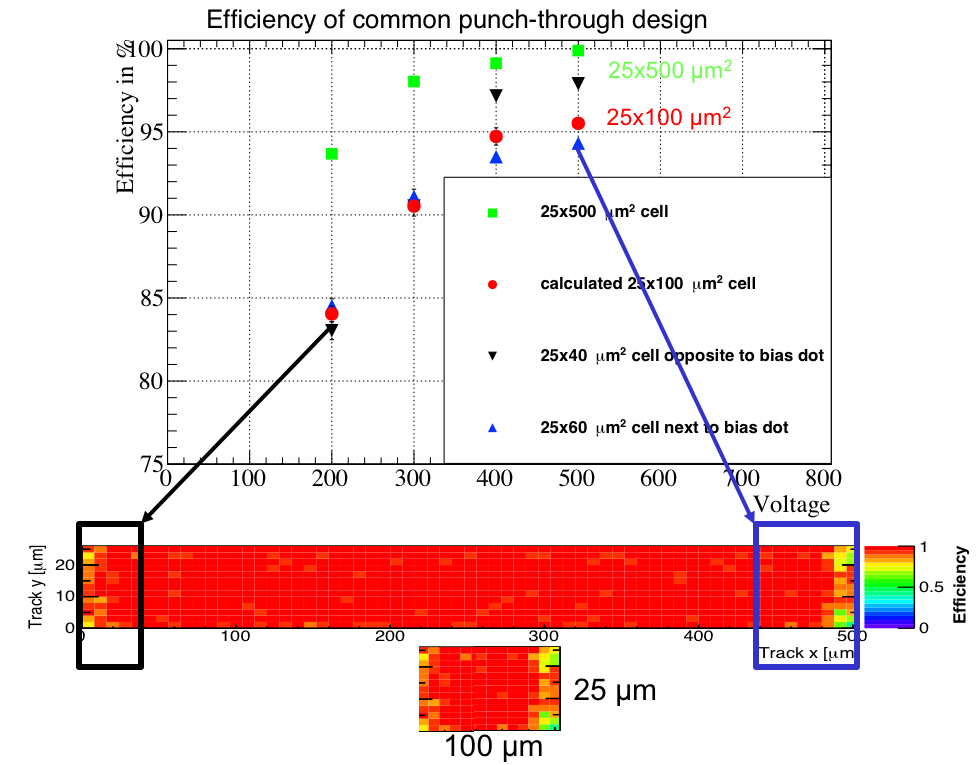}
		\caption[]{Estimated hit efficiency of a 25x100 \SI{}{\micro{m}^2} pixel cell irradiated at \SI{3d15}{\neqcm{}} and compared to the hit efficiency of the larger 25x500 \SI{}{\micro{m}^2} pixel cell. Efficiencies at the edge regions are also shown separately.}
		\label{ptestimation}
	\end{center}
\end{figure}
\section{Performance at high incident angle}
\begin{figure}[h]
	\begin{center}
		\includegraphics[width=0.3\textwidth]{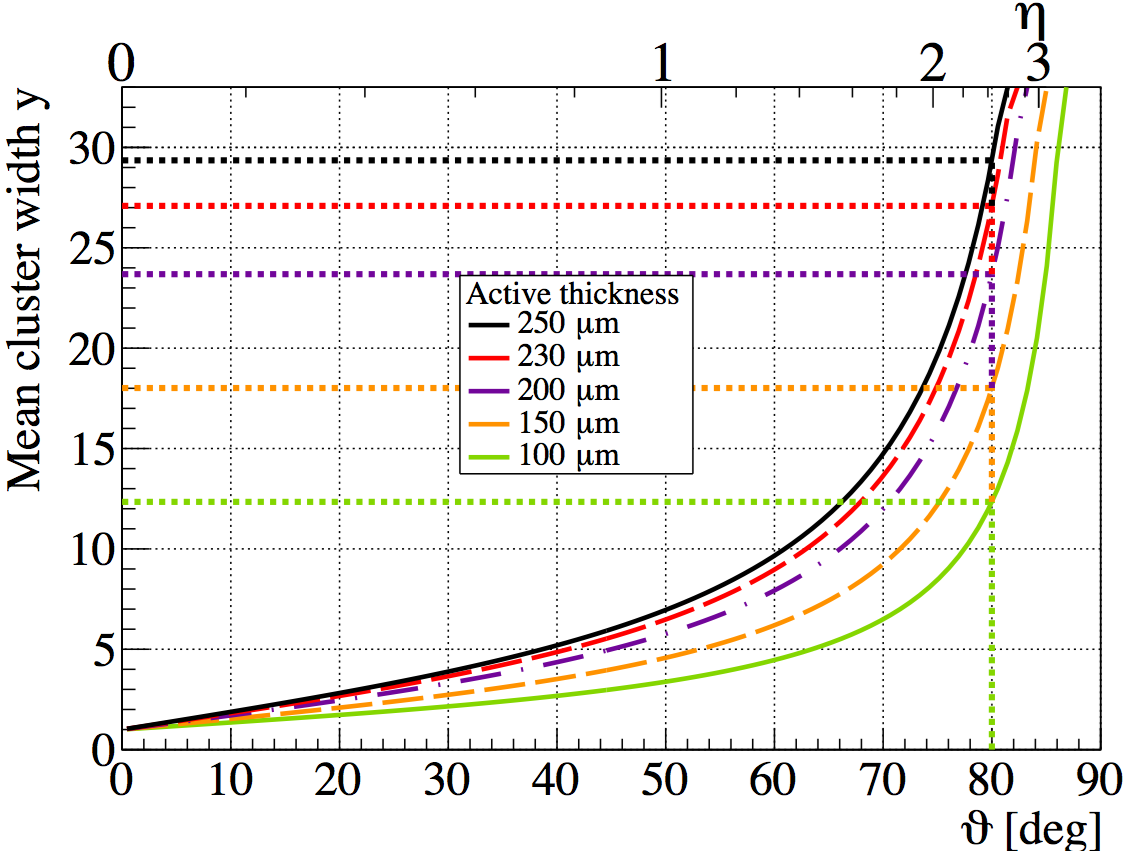}
		\caption[]{Mean cluster width along the short pixel cell side for an FE-I4 module placed at high $\theta$ in the beam, as a function of beam incidence angle. The relationship is also valid for a pixel sensor with a 50x50 \SI{}{\micro{m}^2} cell at high $\eta$ with respect to the beam~\cite{terzo}.}
		\label{cluster}
	\end{center}
\end{figure}

\begin{figure}[h!]
	\begin{center}
		\includegraphics[width=0.28\textwidth]{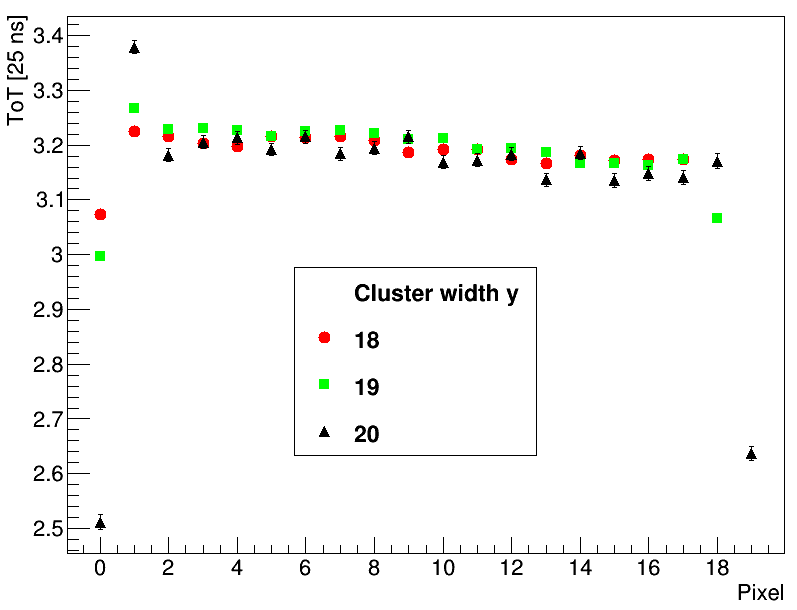}
		\caption[]{Charge collection expressed in units of Time over Threshold, as a function of pixel number for cluster size 18, 19 and 20 obtained with a not irradiated FE-I4 module employing a 200 \SI{}{\micro{m}} thick sensor and tilted by 80$^{\circ}$. The pixel number equal to zero correspond to the sensor back-side and 20 to the sensor front-side.}
		\label{highphi}
	\end{center}
\end{figure}

Since smaller pixel cells are challenging for the tracking in high pseudo-rapidity regions (high $\eta$), the hit efficiency for a cell of 50x50 \SI{}{\micro{m}^2} at $\eta$=2.5 was determined.  FE-I4 modules were placed in the beam at DESY and CERN SPS in such a way that the particles were crossing the pixel along the short side (50 \SI{}{\micro{m}}) at an angle of $\theta$=80$^{\circ}$. Such measurements were previously performed at CERN with a 100 \SI{}{\micro{m}} thick not irradiated module from the VTT production and at DESY with a 200 \SI{}{\micro{m}} thick irradiated module from a CiS production. The cluster size along $\eta$ strongly depends on the sensor thickness: thinner sensors produce smaller clusters and result in a lower pixel occupancy, as shown in Fig.~\ref{cluster}. There was no possibility to reconstruct tracks for these data sets, implying that the analysis needs to be performed using the hit information of the long clusters compatible with the hypothesis a single particle passing through the sensor. Given the particle path of approximately 50 \SI{}{\micro{m}} in each pixel cell, a collected charge of 3100 e was estimated ~\cite{phil}. Therefore, a dedicated tuning was performed with a low target threshold of 1000 e. A not irradiated 200 \SI{}{\micro{m}} thick sensor was recently investigated in a beam test at CERN SPS again at  $\theta$=80$^{\circ}$. An average cluster size of 20 pixels was reconstructed, slightly less than the expected value of 23-24. This was due to a misplacement of 2$^{\circ}$ of the module with respect to the nominal position. 
\\The collected charge is shown in Fig.~\ref{highphi} for the different pixels belonging to the cluster. It is almost constant over the sensor depth, except for the entrance and exit pixels that are not entirely crossed by the track. The obtained hit efficiency of a single pixel for the 200 \SI{}{\micro{m}} thick sensor is 94.9\% and hence it is (4-5)\% lower than the one of 100 \SI{}{\micro{m}} thickness presented in~\cite{anna}. The hit efficiency was calculated with varying assumptions on the allowed number of holes between two pixels inside the cluster, up to a maximum of 10. The single hit inefficiency is the total number of holes divided by the cluster length not including the entrance and exit pixels that are 100\% efficient by construction.
\section{Novel pixel cell design}
\begin{figure}[h]
	\begin{center}
		\includegraphics[width=0.23\textwidth]{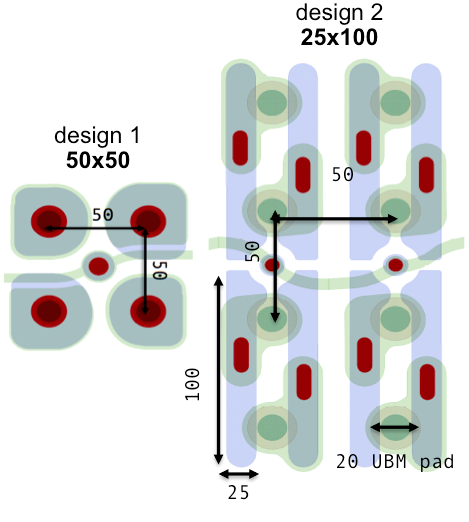}
		\caption[]{Novel pixel design of 50x50 and 25x100 \SI{}{\micro{m}^2} pixel cells combining the novel common punch-through design with the new arrangement of the bias rail being superimposed to the pixel implant as much as possible.}
		\label{newdesign}
	\end{center}
\end{figure}

New pixel cells were designed, compatible with the 50x50 \SI{}{\micro{m}^2} grid of the read-out chip being developed by the RD53 Collaboration: one with the same cell size and an alternative version with 25x100 \SI{}{\micro{m}^2}. For both designs the best performing biasing structures were combined. The common punch-through design was chosen together with the bias rail running as much as possible superimposed to the pixel implant. Both new layouts, displayed in Fig.~\ref{newdesign}, have been implemented in the recent MPG-HLL production described in the next section.
\subsection{Sensor productions at CiS and MPG-HLL}
\begin{figure}[h!]
	\begin{center}
		\includegraphics[width=0.3\textwidth]{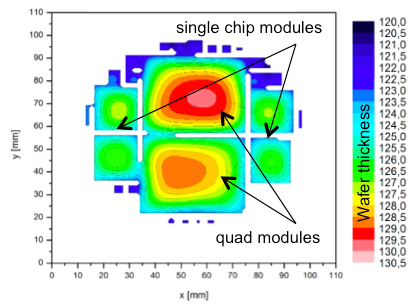}
		\caption[]{Thickness fluctuations within one wafer~\cite{wittig}.}
		\label{thincis}
	\end{center}
\end{figure}

\begin{figure}[h]
	\begin{center}
		\includegraphics[width=0.35\textwidth]{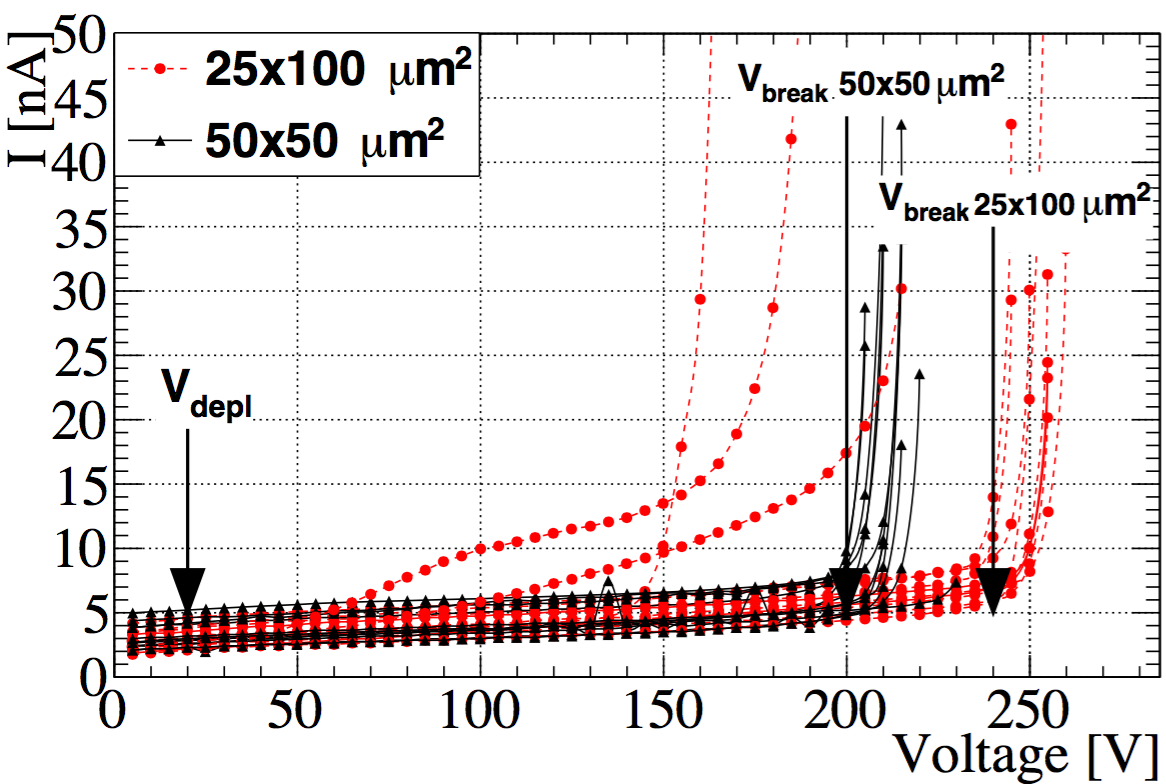}
		\caption[]{Test measurements on one MPG-HLL thin sensor wafer with small pixel cells of 50x50 and 25x100 \SI{}{\micro{m}^2}.}
		\label{ivcurve}
	\end{center}
\end{figure}

At CiS, an innovative method was explored to produce thin sensors without the use of a handle wafer. The technology employs anisotropic KOH etching to create backside cavities in the wafer leaving thicker frames around each single structure. 
\begin{figure}[h]
	\begin{center}
		\includegraphics[width=0.25\textwidth]{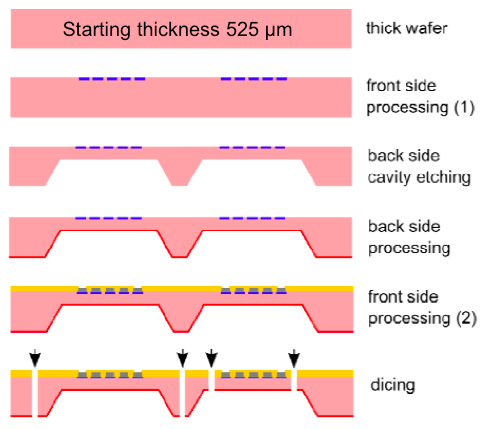}
		\caption[]{Production process flow at CiS technology~\cite{wittig}.}
		\label{thinningcis}
	\end{center}
\end{figure}
\\Fig.~\ref{thinningcis} describes the process flow, starting from the partial front-side processing of 525 \SI{}{\micro{m}} thick wafers and continuing with the cavity etching and the implantation of a p$^+$ layer on the back-side and the p-spray on the front-side followed by a common thermal annealing. The processing finishes with the last metallization and passivation steps on the front-side~\cite{wittig}. The individual thinning steps are shown in Fig.~\ref{thinningcis}. Thanks to the optimized procedure, only small thickness fluctuations in the cavities were measured, up to a maximum value of 10 \SI{}{\micro{m}}, even over the large surface of the quad sensors, see Fig.~\ref{thincis}. A production carried out at MPG-HLL has been recently completed on 6" wafers. The handle wafers will be completely etched away after the Under Bump Bond Metal (UBM) processing at IZM Berlin. 
\\Before this step, the structures have been electrically characterized at wafer level by means of IV curves. Examples of the results are shown in Fig.~\ref{ivcurve} for the RD53 compatible sensors of 150 \SI{}{\micro{m}} thickness. They show very low currents below 10 nA. The breakdown voltages are in the range of 150-250 V and well above the depletion voltage measured to be 15-20 V.
\section{Summary and conclusions}
Pixel modules were investigated for the upgrade of the ATLAS pixel system at HL-LHC. It has been shown that the main inefficiency regions coincide with the biasing structure of the pixel implants on the sensor front side. New punch-through structures with an external bias dot, common for four pixel cells and a new arrangement of the bias rail, have been studied for thicker sensors after an irradiation of \SI{5d15}{\neqcm{}} and compared to the standard design and to previous results at a fluence of \SI{3d15}{\neqcm{}}. It was found that the new structures yield a higher hit efficiency. Novel sensors were designed with a combination of the two biasing structures and implemented in two thin sensor productions at CiS and MPG-HLL with 100-150 \SI{}{\micro{m}} thin sensors, compatible to the new RD53 read-out chip. First measurements of sensors with a pixel cell of 50x50 and 25x100 \SI{}{\micro{m}^2} from the MPG-HLL production have been presented and show leakage currents below 10 nA and breakdown voltages well above the depletion voltages. Those sensors will be interconnected to the prototype chip of the RD53 Collaboration foreseen to be available in 2017. FE-I4 compatible sensors of the same productions will be in the meantime analyzed by means of test-beams to verify the radiation hardness up to fluences of \SI{1d16}{\neqcm{}}.\\
To allow for an investigation of the hit efficiency at high $\eta$ at HL-LHC for the 50x50 \SI{}{\micro{m}^2} pixel cell, FE-I4 modules were tested in beam tests. The cluster properties were analyzed and a good hit efficiency extracted for the single pixels. Thinner sensors are found to be advantageous at high $\eta$ for 50x50 \SI{}{\micro{m}^2} pixel cells given the lower cluster size and hence reduced occupancy expected in the innermost layer. Measurements of the samples after irradiation are planned to be performed with 100-150 \SI{}{\micro{m}} thin sensors in a wider fluence range in the near future.
\section*{Acknowledgement}
This work has been partially performed in the framework of the CERN RD50 Collaboration. The authors would like to thank A. Dierlamm for the irradiation at KIT.
%

\end{document}